# PNO: A Promising Deep-UV Nonlinear Optical Material with Extremely High Second Harmonic Generation Effect

Congwei Xie[a†*], Abudukadi Tudi[b†], Artem R. Oganov[a*]

[a] Skolkovo Institute of Science and Technology, Skolkovo Innovation Center, 121205, Moscow, Russian Federation

[b] CAS Key Laboratory of Functional Materials and Devices for Special Environments, Xinjiang Technical Institute of Physics & Chemistry, CAS; Xinjiang Key Laboratory of Electronic Information Materials and Devices, 40-1 South Beijing Road, Urumqi 830011, China

[†]These authors contributed equally to this work.

In this work, the polar tetrahedron [$PN_2O_2$] was revealed as a new deep-ultraviolet (deep-UV) nonlinear optically active unit. Accordingly, a thermodynamically stable compound (PNO) consisting of the polar [$PN_2O_2$] units was predicted and suggested as a promising candidate of deep-UV nonlinear optical (NLO) material. Compared with other deep-UV materials known to date, PNO possesses the strongest second harmonic generation (SHG) coefficient (about 6 times that of $KH_2PO_4$ (KDP)). Moreover, its three-dimensional connectivity endows it with good mechanical and thermal properties. Therefore, PNO should be a new option for non-π-conjugated deep-UV NLO materials.

Nonlinear optical (NLO) crystals for producing coherent light in the deep-ultraviolet (deep-UV, λ ≤ 200 nm) range are the key materials of solid-state deep-UV lasers.[1] To date, $KBe_2BO_3F_2$ (KBBF) is still the only practical deep-

UV NLO crystal which can directly generate lasers with wavelength shorter than 200 nm by second harmonic generation (SHG), although its further industrial applications are seriously limited by two drawbacks: it contains highly toxic beryllium and is hard to grow its large crystals because of the platelet-like shape of crystals.[2] Therefore, exploring new deep-UV NLO materials is urgently needed and of great interest from both a scientific and industrial standpoint.[3]

A promising deep-UV NLO material should simultaneously satisfy many stringent criteria, such as high deep-UV transparency (band gap $E_g \geq 6.2$ eV), relatively large SHG coefficient (at least comparable to that of classical NLO crystal $KH_2PO_4$ (KDP) 0.39 pm/V), appropriate birefringence for phase-matching (PM) and so on. For the past few decades, materials that could meet these conditions were mainly found in π-conjugated systems, e.g., borates, carbonates and their derivatives, in which the π-conjugated building units ($[BO_3]$, $[B_3O_6]$, $[CO_3]$, etc.) are both beneficial to the increase of SHG coefficient and birefringence.[4]

In addition to π-conjugated systems, non-π-conjugated systems such as silicates, phosphates and sulfates, have also been considered as potential candidates for deep-UV NLO materials, where tetrahedral groups $[XO_4]$ (X = P, S, Si) are favorable with deep-UV transparency.[5] Many NLO-active phosphates, sulfates and silicates with high deep-UV transparency and SHG coefficient have been discovered. Unfortunately, none of them possess sufficient birefringence to satisfy the PM conditions in the deep-UV region because of weak optical anisotropy of nonpolar $[XO_4]$ tetrahedral groups in their structures. Thus, regarding non-π-conjugated NLO materials, the greatest challenge lies in how

to achieve enough birefringence to realize PM in the deep-UV region.[6]

The common strategy to improve birefringence is to introduce transition metal cations with $d^0$ or $d^{10}$ electronic configuration or $ns^2$ lone pair cations.[7-8] However, in most cases, this causes a heavy red-shift of the absorption edge and thus does not work to design deep-UV materials. Recently, polar polyhedra in which the central positive ions are bonded to two anionic ligands (e.g., O, F, NH$_2$, etc.) were proposed as good functional structural units (FSUs) for designing deep-UV NLO materials.[3, 9-11] Among them, polar tetrahedra [BO$_{4-x}$F$_x$], [PO$_{4-x}$F$_x$], [SO$_{4-x}$F$_x$] and [SO$_{4-x}$(NH$_2$)$_x$] ($x$ = 1, 2) have been found to possess large polarizability anisotropy and hyperpolarizability and thus have advantages on enhancing birefringence and SHG coefficients.[3, 9-11] Accordingly, several fluoroborates, fluorophosphates and sulfamides have been theoretically reported and/or experimentally confirmed as deep-UV NLO materials.[3, 9b, 10a, 11]

We noticed that nitrogen can also be used to substitute oxygen in [XO$_4$] units. It has been reported that the introduction of nitrogen into [SiO$_4$] unit can form polar [SiNO$_3$], [SiN$_2$O$_2$], and [SiN$_3$O] tetrahedra with large polarizability anisotropy and high hyperpolarizability, and oxonitridosilicates containing these polar tetrahedra have been identified as new candidate systems to explore new UV NLO materials.[12]

In this work, our theoretical calculations showed that the polar [PN$_2$O$_2$] tetrahedron has large polarizability anisotropy and hyperpolarizability, relatively large highest occupied molecular orbital (HOMO)-lowest unoccupied molecular orbital (LUMO) gap and optical gap, indicating that [PN$_2$O$_2$] tetrahedron should be a good deep-UV NLO-active unit. To confirm this, a thermodynamically stable noncentrosymmetric (NCS) compound (PNO) consisting of polar [PN$_2$O$_2$] tetrahedra was predicted for evaluating its optical properties. The present theoretical study showed that PNO has a short deep-UV cut-off edge about 188 nm and PM wavelength less than 190 nm. In addition, benefiting from both the intrinsic excellent optical properties of [PN$_2$O$_2$] tetrahedra and their ordered arrangement, PNO achieves the strongest SHG coefficient (about 6 times that of KDP) among the whole deep-UV NLO materials. Similar to [PN$_2$O$_2$] tetrahedron, [PNO$_3$] is also newly suggested to be a potential DUV-NLO unit. However, we didn't predict any stable ternary P$_x$N$_y$O$_z$ compounds containing [PO$_3$N] tetrahedron.

We have evaluated the basic electronic and optical properties for a serials of Si and P based tetrahedral anionic groups: [SiO$_{4-x}$F$_x$] ($x$ = 0,1,2,3), [SiN$_x$O$_{4-x}$] ($x$ = 1,2,3,4), [PO$_{4-x}$F$_x$] ($x$ = 0,1,2) and [PN$_x$O$_{4-x}$] ($x$ = 1,2,3,4). We first performed geometry optimization and then investigated their electronic structure and optical properties based on DFT calculations. As shown in Fig. 1 and listed in Table S1 in the supporting information, the polar [PN$_2$O$_2$] and [PNO$_3$] units exhibit high polarizability anisotropy of 12.55 and 8.80, respectively. As a comparison, the polarizability anisotropies of well-known deep-UV NLO-active units [PO$_2$F$_2$] and [PO$_3$F] calculated by using the same computational settings is 9.02 and 6.86, respectively. The [PN$_2$O$_2$] and [PNO$_3$] units also have larger hyperpolarizabilities than those of [PO$_2$F$_2$] and [PO$_3$F] units. Although the substitution of O by N will reduce the HOMO–LUMO gap and optical gap, [PN$_2$O$_2$] and [PNO$_3$] units still have relatively large HOMO-LUMO gaps (> 5 eV) and optical gaps (> 4 eV). We here note that, in P$_x$N$_y$O$_z$ compounds, N atoms will be coordinated with more than one P atom, which can eliminate the non-bonding electrons and thus helpful to enhance the band gap. Thus, similar to [PO$_3$F] and [PO$_2$F$_2$] units, [PN$_2$O$_2$] and [PNO$_3$] units can also balance the HOMO-LUMO gap, hyperpolarizabilities and polarizability anisotropies, suggesting that they are good deep-UV NLO active units.

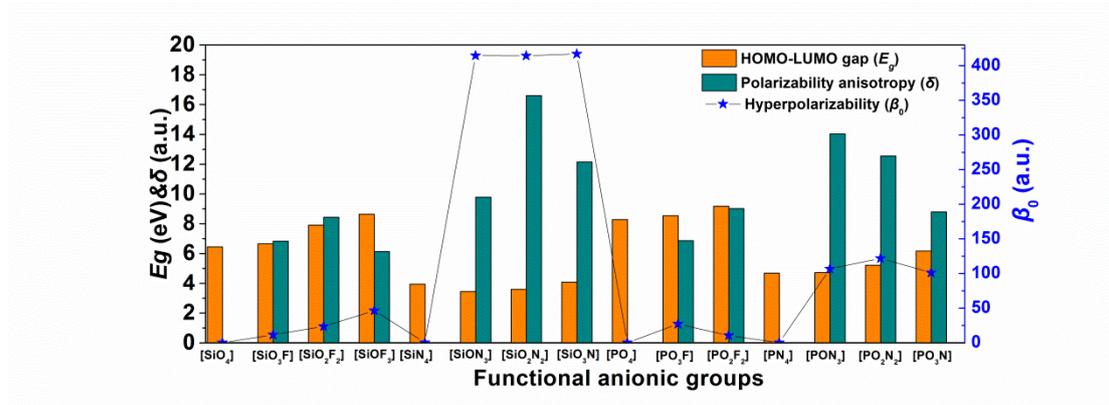

**Fig. 1**. Computed HOMO-LUMO gaps (Eg, in eV), polarizability anisotropies (δ, in a.u., 1 a.u. = 1.6488 ×10$^{-41}$ C$^2$m$^2$J$^{-1}$) and the mean static first hyperpolarizability (β$_0$, in a.u., 1 a.u. = 3.20636 ×10$^{-53}$ C$^3$m$^3$J$^{-2}$) of [SiO$_{4-x}$F$_x$] (x = 0,1,2,3), [SiO$_{4-x}$N$_x$] (x = 1,2,3,4), [PO$_{4-x}$F$_x$] (x = 0,1,2) and [PO$_{4-x}$N$_x$] (x = 1,2,3,4) tetrahedral anionic groups.

To explore potential deep-UV P$_x$N$_y$O$_z$ compounds, using evolutionary algorithm as implemented in the USPEX code[13-15], we performed variable composition structure searches for the P-N-O system with up to 36 atoms in the unit cell at zero temperature and ambient pressure. Based on total energy computed for all the predicted structures, we constructed the phase diagram for the P-N-O system. As shown in Fig. 2, along with the experimentally reported P$_4$N$_6$O[16], we can find that PNO is also thermodynamically stable. Luckily, the lowest-energy PNO structure (PNO-I) is a non-centrosymmetric structure, see Table S2 in the supporting information. In addition to the stable non-centrosymmetric structure, we have also predicted one competitive metastable non-centrosymmetric structure (PNO-II); its energy is slightly higher (by 1 meV/atom) than that of PNO-I. We here note that, both PNO-I and PNO-II can be also produced by structural analogy technique, as reported by Materials Project[17]. For PNO-II, along with other metastable PNO structures predicted by USPEX, their details of structural and thermodynamic information are also listed in Table S2 in the supporting information.

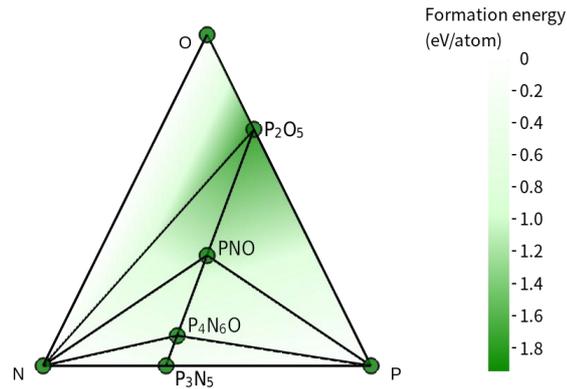

**Fig. 2.** Phase diagram of P-N-O system – only stable compounds are shown and they are denoted by circles

With fully relaxed structures, we computed phonon dispersion curves and elastic constants for confirming mechanical and dynamical stability of PNO-I and PNO-II. It was found that both these two PNO structures are mechanically and dynamically stable (see Fig. S2 and Table S3 in the supporting information). Crystal structures of PNO-I and PNO-II are shown in Fig. 3. The asymmetric unit of PNO-I contains one unique P atom, one unique O atom and one unique N atom and each P atoms are connected to two O atoms and two N atoms forming the FSU [PN2O2] tetrahedra (see Fig. 3(a)). PNO-I belongs to $I2_12_12_1$ space group (see Fig. 3(b)), the corner-sharing [PN2O2] tetrahedra form a three-dimensional framework structure (see Fig. 3(c)), derived from tridymite (see Fig. 3(d)). PNO-II also has one unique P atom, one unique O atom and one unique N atom and has the same FSU found in PNO-I. PNO-II belongs to *Cc* space group (see Fig. 3(e)) and also has a 3D framework structure derived from cristobalite (see Fig. 3(f) and 3(g)). We here note that this 3D-connectivity nature of PNO is helpful to yield superior properties such as good mechanical strength and thermal conductivity (see Table S3 in the supporting information) and thus may benefit to its further applications.

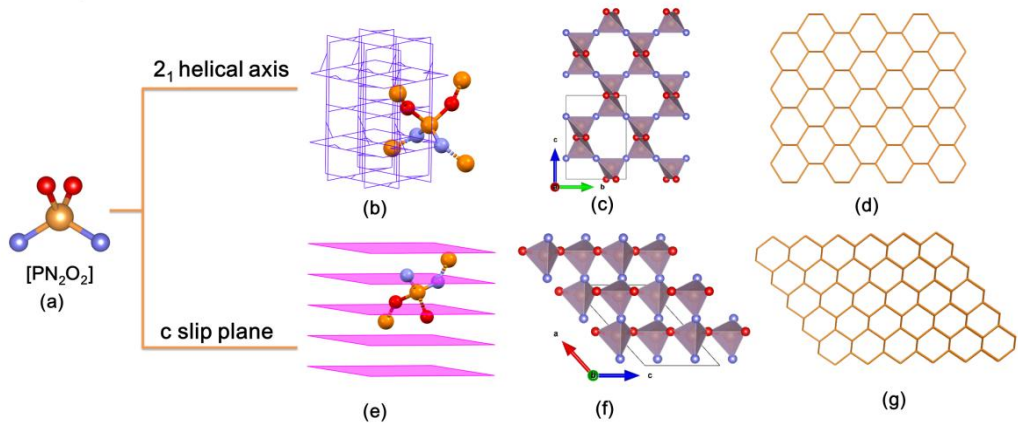

**Fig. 3**. (a) [PN$_2$O$_2$] unit, (b) 2$_1$ helical axis in a, b, c direction, (c) 3D structure of PNO-I, (d) topology of PNO-I, (e) c slip plane in c direction, (f) 3D structure of PNO-II, (g) topology of PNO-II

The NLO properties of PNO-I and PNO-II are shown in Table 1. Both predicted structures have relatively wide band gaps greater than 6.2 eV, which demonstrates their ability to achieve deep-UV transmission. They both have large birefringence (> 0.151 @ 1064 nm) beneficial to realize PM. The PM wavelengths of PNO-I and PNO-II are lower than 190 nm (see Fig. S3 in the supporting information), and the static SHG coefficients of these two structures are large. The SHG coefficients and PM wavelengths of PNO structures along with those of other well-known deep-UV materials are shown in Fig. S5. We can find that PNO-I and PNO-II have very large SHG coefficients (about 6-9 times that of KDP). Compared with other compounds containing non-π-conjugated groups and π-conjugated groups, PNO has the largest SHG coefficient.[18]

**Table 1**. Calculated band gaps by Hybrid HSE06 functional ($E_g$-HSE, in eV), the independent SHG coefficients at zero frequency (SHG, in pm/V), birefringence (Δn@1064 nm), and the shortest phase-matching wavelengths ($\lambda_{PM}$, in nm) for PNO-I and PON-II.

| Compound | | Space groups | $E_g$-HSE | SHG | Δn | $\lambda_{PM}$ |
|---|---|---|---|---|---|---|
| PNO-I | Cal | $I2_12_12_1$ | 6.61 | $d_{36} = 2.33$ | 0.151 | < 190 |
| PNO-II | Cal | $Cc$ | 6.63 | $d_{11} = -3.44$; $d_{31} = 0.89$ | 0.174 | < 190 |
| | | | | $d_{26} = 2.16$; $d_{35} = 0.68$ | | |
| | | | | $d_{32} = 1.01$; $d_{33} = -0.87$ | | |

In order to resolve the source of the excellent NLO properties of PNO, we performed structural performance analysis using first-principles calculations for PNO-I. As shown in Fig. 4(a), PNO-I has direct band gap with HSE06-value of 6.61 eV. The valence bands (VBs) are mainly composed of N-2p states, O-2p states and P-3p states, and the conduction bands (CBs) are mainly derived from P-3p states (Fig. 4(b)). The SHG density and band-resolved methods[19] were used to analyze the largest SHG coefficient PNO-I (Fig. 4(b)-4(d) and Fig. S6 in the supporting information).

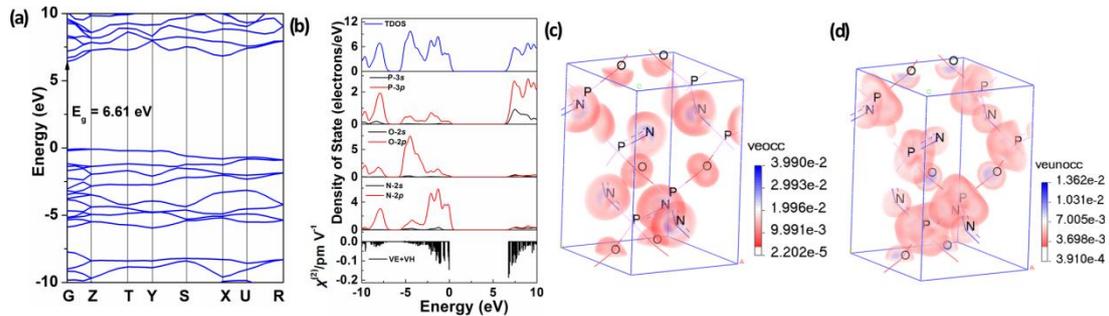

**Fig. 4**. Band structure, (b) density of states band-resolved NLO susceptibility χ(2), the SHG-density of occupied (c) and unoccupied (d) states

The virtual-electron (VE) and the virtual-hole (VH) transitions account for 72% and 28% of the SHG effect, respectively. The VE process is shown in Fig. 4(c) and 4(d). It can be found that the occupied states mainly originate from O-2p and N-2p orbitals, and the unoccupied states are mainly from O-2p, N-2p states and P-3p states. The VH process is shown in Fig. S5 in the supporting information. The VH occupied states are mainly the contribution of O-2p and N-2p orbitals, and the unoccupied states are mainly the contribution of O-2p, N-2p states and P-3p states. Those results are consistent with the band-resolved analysis (see Fig. 4(b)). It is shown that N-2p and O-2p orbitals make the major contribution for the SHG response in the valence band (-2.5 ~ 0 eV) region. In the conduction bands, the P-3p, N-2p and O-2p orbital in the 6.61 ~ 10 eV region makes a major contribution to the SHG response.

In summary, we suggest the polar [$PN_2O_2$] tetrahedron as a new FSU for designing non-π-conjugated deep-UV NLO materials because of its large hyperpolarizabilities, polarizability anisotropies and suitable HOMO-LUMO gap. To confirm this, we performed crystal structure search for P-N-O system to explore potential deep-UV structure. We discovered a thermodynamically stable NCS PNO structure and a competing metastable one. Both of them have large band gap 6.61~6.63 eV, large birefringence 0.151~0.174 @1064 nm and large SHG responses about 6~9 times of KDP. As compared with other deep-UV materials assembled from non-π-conjugated groups and even those π-conjugated groups, PNO has the largest SHG coefficients.

## Author Contributions

C.W.X. and A. R. O. conceived and designed the research. C.W.X. and A. T. performed all research work. C.W.X. and A. T. performed the theoretical analysis and wrote the paper. All the authors discussed the results and commented on the manuscript.

## Conflicts of interest

The authors declare no conflict of interest.

## Acknowledgements

This work was supported by the Russian Science Foundation (grant 19-72-30043). The authors also acknowledge the CAMD Laboratory for the allocation of computing time on their machines.

## Notes and references